\documentclass[aps,twocolumn]{revtex4-2}
\usepackage{amsmath,amssymb,amsfonts}
\usepackage{algorithmic}
\usepackage{graphicx}
\usepackage{textcomp}
\usepackage{xcolor}
\usepackage{etoolbox} % toggles
\usepackage[version=4]{mhchem}

\usepackage{breakcites}

\newcommand{\D}{\mathrm{d}} 		% for derivative
\newcommand{\E}{\mathrm{e}} 		% for exponential function
\newcommand{\I}{\mathrm{i}} 			% for imaginary unit
\newcommand\arcdeg{\mbox{$^\circ$}} % degrees symbol
\renewcommand{\epsilon}{\varepsilon}

\begin{document}
\title{Energy landscape interpretation of universal linearly increasing absorption with frequency}
%Universality of linearly increasing absorption with frequency}
%\title{Universal linear  frequency absorption  due to uniform activation energy distribution}
% linear dependence of absorption on frequency
\author{Sverre Holm and Joakim Bergli} %,\\ Department of Physics, University of Oslo}
\affiliation{Department of Physics, University of Oslo}
%\maketitle % for article
%%%%%%%%%%%%%%%%%%%%
\begin{abstract}
%Absorption of elastic waves in complex media often depends on frequency in a linear way for both longitudinal and shear waves. This universal property occurs in  media such as

Absorption of elastic waves in complex media is commonly found to increase linearly with frequency, for both longitudinal and shear waves. This ubiquitous property is observed in media such as rocks, unconsolidated sediments, and human tissue.  
Absorption is due to relaxation processes at the level of atomic scales and up to the sub-micron scale of biological materials. The effect of these processes is usually expressed as an integral over relaxation frequencies or relaxation times. Here we argue that these processes are thermally activated. 
Unusual for ultrasonics and seismics, we can therefore transform the expression for absorption from the frequency or time domains to an integral over an activation energy landscape weighted by an energy distribution. The universal power-law property surprisingly corresponds to a flat activation energy landscape. This is the solution which maximizes entropy or randomness. Therefore the linearly increasing absorption corresponds to the energy landscape with the fewest possible constraints.
\end{abstract}

\maketitle % for APS

\section{Introduction}
%%%%%%%%%%%%%%%%%%%%%%%%%%%%%%%%%%%%%%%%%%%

The amplitude of elastic waves, i.e.~both compressional and shear waves, undergoes attenuation which often follows a power-law in frequency, $\omega$:
\begin{equation}
|u(x, \omega)| \propto \E^{-\alpha(\omega) x}, \enspace \alpha(\omega) = \alpha_0\omega^{y},
\label{eq:PowerLawabsorption}
\end{equation}
where ${u}$ is particle velocity and $x$ is distance. Attenuation, $\alpha$, has unit m$^{-1}$ or Nepers/m and is also called inverse mean free path. The two mechanisms for attenuation are viscoelastic absorption leading to heating, and  scattering of energy from inhomogeneities. This paper is only concerned with mechanisms for absorption, although medium properties may also be inferred from the scattering \cite{annio2024making}. In complex media absorption often increases linearly with frequency, i.e.~$y=1$ \cite[Sect.~5.1]{holm2019waves}, as the many examples that follow demonstrate.

In seismology, many nuclear-explosion and earthquake data sets in the  range $10^{-3}$ to 10 Hz appear to have a constant quality factor, $Q$, for shear waves \cite{morozov2010causes}. The unit-less inverse Q or attenuation per wavelength, (also called specific attenuation or internal friction), is \cite[Sect.~2.3]{holm2019waves}:
\begin{equation}
Q^{-1} =  \frac{\alpha(\omega) \lambda(\omega)}{\pi} 
= 2\frac{\alpha(\omega) c(\omega)}{\omega},
\label{eq:InverseQ}
\end{equation}
where $\lambda$ is wavelength. When there is power-law absorption and dispersion is small, $c(\omega) \approx  c_0$ \cite{waters2000applicability}, the inverse Q will be proportional to $\omega^{y-1}$. Therefore,
linearly increasing absorption results in constant-Q behavior. 

Constant-Q behavior is especially apparent after correction for bias due to an additive constant  in the expression for $\alpha(\omega)$, which may be due to geometrical spreading, (de-)focusing, or scattering \cite{morozov2010causes}.
Constant-Q behavior is also reported for seismic reflection data from vertical wells \cite{gurevich2015frequency} and such behavior is also  common  in seismic survey data up to about 100 Hz  \cite{campbell2009estimates}.

It is also recognized that $Q$ is nearly frequency-independent over one to two decades of frequency in many solids \cite{Knopoff1964Q}.  This property was noted for metals and nonmetals,  for both compressional and shear waves, and for frequencies in the Hz, kHz, and MHz range. 

Unconsolidated sub-bottom sediments represent a very different medium, but the same linearly increasing absorption of compressional waves is  observed above about 2 kHz  \cite{williams2002comparison}. Below that frequency the exponent is closer to two. Even in this case the value for the exponent, $y$, will depend on whether shear-wave mode conversion in the form of a constant term 
in the expression for $\alpha(\omega)$ 
is compensated for \cite{carey2008exponent}.

One of the best studied fields is compressional waves in the MHz range in medical ultrasound, %\cite{wells1975absorption},
 \cite{kadaba1980attenuation}. A recent review says that  an ``early consensus emerged that a power law fit near 1 was adequate for absorption models of soft tissues'' \cite{parker2022power}. As in the case of seismology, care is needed in interpreting measurements as  defocusing due to phase aberrations may lead to an overestimation of absorption if not compensated for \cite{marcus1975problems}. 

The medical ultrasound field has provided insight into the spatial scale where absorption takes place. Grinding liver tissue hardly changes the absorption, so it is apparent that the mechanisms operate "on a level of organization smaller than that defined by cells, cell nuclei, and mitochondria", i.e., size less than about 3 microns  \cite{Pauly1971}. Absorption in canned evaporated milk also  follows a linear frequency law   \cite{carstensen1954measurement}, indicating that casein micelles at the sub-micron size level play an important role. 
Similarly, absorption of compressional waves in blood in the 0.8-3 MHz range was due to the presence of proteins \cite{carstensen1953determination}.

%A related phenomenon is how cells or tissue react to shear excitation. The response often varies with $\omega^\beta$, where $\beta$ is in the range 0.1 to 0.5 \cite{kollmannsberger2009active}, \cite{parker2019towards}. According to results cited in the Supplement, this corresponds to an absorption that varies with $y=\beta+1$ and therefore $y$ in this case will fall in the range 1.1 to 1.5.

The property that the exponent $y$ is near unity over a wide range of  materials is nearly universal, both for compressional and shear waves, as noted in e.g.~\cite{Szabo00}. The exponent may in some cases be larger than unity, but never above 2, the viscous case, and rarely below 1. The temperature range of interest is approximately 0 to 40$^{\circ}$~C.

This paper is concerned with viscoelastic absorption, but there are other mechanisms which will not be discussed such as friction in cracks in rocks, fluid flow in micropores in consolidated sediments or bone \cite{meziere2014measurements}, 
squirt flow in unconsolidated sediments \cite{chotiros2014shear}, 
and nonlinearity.

An early study of absorption in erythrocytes suggested that  ``chemical or structural relaxation processes are probably responsible for the attenuation'' \cite{kremkau1973macromolecular}. It has also been proposed that in polymer-like materials, thermal energy causes continuous change in the 
``interactions between macromolecules'' \cite{Szabo00}. %In cell rheology, the theory of soft glassy materials has been shown to be applicable \cite{kollmannsberger2009active}.

%We also notice that chemical and structural relaxation in e.g., seawater are thermally activated processes. Thermal activation over potential barriers of structural defects also plays an important role in e.g. glasses and rocks. This gives the motivation to transform the multiple relaxation formulations from the usual frequency and time domains into the energy domain. This will provide new insights for the case of elastic wave propagation in acoustics and seismics.  Surprisingly, a simple uniform distribution of activation energies corresponds to the universal linearly increasing absorption.

Here we claim that it can be justified to transform the multiple relaxation formulation  into the energy domain for many of the materials just listed. This is a transformation into an energy landscape which is common for describing atomic and molecular clusters, glasses, and proteins. 

We follow \cite[Chap.~1]{wales2004energy} and define an energy landscape as either a
potential energy surface or a free energy surface. Of these, the
potential energy surface is the more fundamental as it gives the
potential energy as a function of atomic or molecular coordinates. In
complex media, this would be a function of many variables, which is
unrealistic to know in detail. One then will replace it by a thermally
averaged free energy surface which is described in terms of a smaller
number of effective degrees of freedom. The distinction will not be
important for us in the following, as we are only assuming certain
statistical properties of the energy landscape, while not relating
them to the microscopic structure.

Surprisingly, a simple uniform distribution of activation energies, i.e. maximal randomness, corresponds to  universal linearly increasing absorption.

The paper starts with multiple relaxation formulations in frequency and time of power-law absorption. Then we introduce the Arrhenius law  and use it to transform the relaxation formulation into the energy  domain with an accompanying energy distribution. The question of whether energy landscapes also describe biological materials in then discussed. The paper ends with a discussion of non-Arrhenius behavior, the effect of band-limited power law absorption, and a comparison with a similar result as ours found in a completely different way in the field of soft glassy materials.

\section{Background: Multiple relaxation}

A single relaxation process  is characterized by a relaxation frequency, $\Omega$, or a relaxation time $\tau=1/\Omega$, and an absorption given by:  
\begin{equation}
   \alpha(\omega) = A \frac{\Omega \omega^2}{\omega^2+\Omega^2}, 
   \label{eq:Relaxation}
\end{equation}
where $A$ is a constant with unit inverse velocity. 
Attenuation increases with $\omega^2$ well below the relaxation frequency and is constant well above it. This expression can for instance be found from structural relaxation \cite{hall1948origin}, and from chemical relaxation \cite{verma1959ultrasonic}.

%
%\subsection{Frequency and time descriptions}
	\label{sec:Multiple}
	%%%%%%%%%%%%%%%%%%%%%%%%%%%%%%%%%%%%%%%%%%%
 
In a complex medium there are many elementary relaxation processes over a large spread of relaxation frequencies and absorption is:
\begin{equation} 
\alpha(\omega) = A_0 \; \omega^2 \int_0^\infty    \frac{g_\Omega(\Omega) \,\Omega}{\omega^2+ \Omega^2} \, \D \Omega.
\label{eq:MultipleRelaxation}
\end{equation}
Following the terminology of  \cite{pierce2021acoustic}, $A_0=2/c_0$, where $c_0$ is the equilibrium sound speed in the limit of zero frequency. 
The weighting $g_\Omega(\Omega)$ in the relaxation integral has the form of a probability distribution function. % when normalized. 
The particular distribution given by 
\begin{equation}
g_\Omega(\Omega) = K_y \; \Omega^{y-2},
\label{eq:pdfRelaxationFrequencies} 
\end{equation}
results in the power-law absorption of \eqref{eq:PowerLawabsorption} \cite[Sect.~3.241.2]{gradshteyn2014table}, \cite{Nasholm2011}, with a normalization factor $K_y$ with unit [s$^{y-1}$]. This result was in fact already found in 1959 \cite[Fig.~8]{carstensen1959acoustic}. This formulation does, however, not provide much insight to motivate why $g_\Omega(\Omega)$ should  follow this particular power-law relation.

The integral of \eqref{eq:MultipleRelaxation} can be transformed to be over relaxation times, $\tau = 1/\Omega$, by letting $g_\Omega(\Omega) = g_\tau(\tau) \; |\D \tau /  \D \Omega |$:
\begin{equation} 
\alpha(\omega)= A_0 \; \omega^2 \int_0^\infty  \frac{  g_\tau (\tau) \; \tau}{1+ \omega^2 \tau^2} \D \tau, 
\label{eq:MultipleRelaxationTime}
\end{equation}
where the particular distribution 
\begin{equation}
g_\tau(\tau) = K_y  \tau^{-y},
\label{eq:pdfRelaxationTimes} 
\end{equation}
will lead to the desired power-law absorption \cite{pierce2021acoustic}. This formulation may be easier to interpret as relaxation times may be related to length scales, $l=c \tau$, by means of the speed of propagation, $c$. 
A possible explanation for relaxation behavior can therefore be that there is a hierarchy of geometrical structures, from large $l$ to small $l$. 

This interpretation has in particular been attempted in 
cell biomechanics which is a field where power-law behavior of the shear modulus over about five decades of frequency is well documented.  Geometrical structures that may be invoked are the cell membrane, the actin cortex, the cytoskeleton etc, all at different length scales. It has been concluded however, that there are not enough cell components in a hierarchy from the largest to the smallest length scales to account for the observed  power-law behavior, see \cite{fabry2001scaling, kollmannsberger2009active, kollmannsberger2011linear} for the detailed argument.  It seems therefore as if there is limited physical insight to gain even from the formulation of \eqref{eq:MultipleRelaxationTime}.

%This is hard to justify in e.g. ultrasonics, but cell biomechanics may offer some insights, as it is a field where power-law behavior of the shear modulus over about five decades of frequency is well documented. Geometrical structures that may be invoked are the cell membrane, the actin cortex, the cytoskeleton etc. 
%However, even in that field there seems not to be enough cell components in a hierarchy from the largest to the smallest to account for the observed  power-law behavior \cite{fabry2001scaling, kollmannsberger2009active, kollmannsberger2011linear}. There is therefore limited physical insight to gain even from the formulation of \eqref{eq:MultipleRelaxationTime}.

\section{Thermally activated relaxation}

In acoustics one is usually content with the descriptions of \eqref{eq:MultipleRelaxation} and \eqref{eq:MultipleRelaxationTime}. The limitation, as noted, is that there is little insight to gain into why the relaxation processes are "organized" to give power-law characteristics. 

A closer look at the common mechanisms for intrinsic absorption is therefore warranted. In acoustics, they are, according to \cite[Chap.~8]{kinsler1999fundamentals}, \cite[Sect.~4.1]{holm2019waves}, viscosity, molecular thermal relaxation, heat conduction in monatomic gases,  structural relaxation, and chemical relaxation. Viscous absorption and molecular thermal relaxation due to oxygen and nitrogen dominate in air, and the three first mechanisms therefore primarily characterize  absorption in gases. 
Only the two last ones are relevant to properties of fluids and solids, our main interest in this paper.
We therefore start by reviewing the well-established model for seawater, as an example of a medium where structural (also called segmental) relaxation and chemical relaxation dominate.

\subsection{The Arrhenius law}

Absorption in seawater has three main components of the form of \eqref{eq:Relaxation}. 
Two of them are due to B(OH)$_3$ and MgSO$_4$ which both contribute two-state chemical equilibrium reactions. Their relaxation frequencies are in the kHz and tens of kHz range respectively.
Relaxation time in these reactions is related to activation energy, $E_a$, according to the Arrhenius relation \cite{verma1959ultrasonic}: %schulkin1978low}:
\begin{equation}
\tau = \tau_0 \E^{E_a/k_BT}, %\enspace \Omega =  \Omega_0 \E^{-V/k_BT}
\label{eq:Arrhenius}
\end {equation}
where $T$ is absolute temperature, $k_B$ is  Boltzmann's constant, and $\tau_0$ characterizes the smallest time scale probed. 
It should be noted that in underwater acoustics, chemical relaxation is usually parameterized with temperature in Celsius rather than Kelvin \cite{Ainslie1998}, obscuring the fact that these processes are thermally activated. 

The most important contribution to absorption in water is from structural relaxation of H$_2$O molecules. In distilled water a broken-down  structure of clusters is dynamically  changed by an incoming sound wave and relaxation takes place as  clusters of different sizes interact. 
Based on a two-state energy model for H$_2$O molecules, the Arrhenius law  describes the transition rates \cite{hall1948origin}. The relaxation frequency is in the THz range.

The Arrhenius law points to an activation energy landscape interpretation of chemical and structural relaxation taking place at the molecular or molecular cluster level, i.e. at the nanometer scale.
For now, that only covers some of the cases mentioned in the Introduction, but  let us anyway pursue the consequences of this view point.

%?? Energy landscape definition \cite[Chap.~1]{wales2004energy}
%\begin{itemize}
%\item Potential energy surfaces: function of all relevant atomic or molecular coordinates, more fundamental
%\item Free energy surfaces: first in spin glass theory, averaged PES, a broader view, temp. dependent
%\end{itemize}

\subsection{Transformation of relaxation integral}

The validity of the Arrhenius relation
is a key insight that allows us to transform the previous relaxation integrals.
Equation \eqref{eq:MultipleRelaxationTime} can be transformed by using \eqref{eq:Arrhenius} in combination with $g_\tau(\tau) = g_E(E_a) |\D E_a / \D \tau |$:
\begin{equation}
\alpha (\omega) = {A_0}\, \omega^2 \int_0^\infty  \frac{g_E(E_a) \tau(E_a)}{1+\omega^2 \tau^2(E_a)} \D E_a,
\label{eq:MultipleRelaxationEnergy}
\end{equation}
where  $g_E(E_a)$ is an energy distribution function. In this way, the Arrhenius law provides a link that allows us to transform the distribution of relaxation rates  to a distribution of activation energies. These can then be interpreted in terms of the energy landscape.

%It should be noted that activation energy and its distribution function are physical parameters that lend themselves to interpretation much more than distributions of relaxation frequencies or relaxation times. 

\subsection{Glass}

It was the use of \eqref{eq:MultipleRelaxationEnergy} for describing the effect of structural defects in glassy media which inspired the work reported here. A glass can be considered to be frozen in an energy basin with many local minima. The height of the barriers between them, and the energy difference between the minima will determine properties \cite{buchenau2022sound}. Structural relaxation takes place due to perturbations of the landscape 
and this is the main cause of absorption. We are mainly concerned with %normal temperatures 
typical terrestrial surface temperatures (about 270 - 310 K), and then tunneling 
 \cite{galperin1989localized} 
can be neglected and the classical model with the thermal activation rate of 
\eqref{eq:Arrhenius} describes the relaxation  \cite{anderson1955ultrasonic}, \cite{jackle1976elastic}, \cite{buchenau2001mechanical}

This describes what happens in structural glasses, %~\cite{phillips1987two}, 
but is also applicable to rocks \cite{carcione2020seismic}, where most minerals consist of crystalline grains with amorphous grain boundaries with structure similar to glasses. 

In the glass field, the main interest is absorption as a function of temperature, and secondarily as a function of frequency. The model of \eqref{eq:MultipleRelaxationEnergy} may either be used to find a distribution of activation energies, $g_E(E)$ that fits experimental data for absorption, or aspects of the distribution may be found from material properties. Common distributions are a gaussian \cite{hunklinger1981acoustic}, a gaussian weighted by a power law \cite {vacher2005anharmonic}, and an exponential \cite{gilroy1981asymmetric}:
\begin{equation}
g_E(E_a) =\frac{1}{E_0} \E^{\frac{-E_a}{E_0}},
\label{eq:GlassExp}
\end{equation}
where $E_0$ is an energy related to the glass temperature. %Uniform distributions are seldom encountered in the glass literature.

The property that absorption follows a power-law as in \eqref{eq:PowerLawabsorption} is not common to see in the description of glassy media, even though \eqref{eq:MultipleRelaxationEnergy} is used extensively. One example that has power-law characteristics is found in  \cite{anderson1955ultrasonic}, where they found an inverse Q that increases linearly with frequency, i.e. $y \approx 2$ in \eqref{eq:InverseQ}. This was found to correspond to a distribution of energies that falls off for higher energies,  in general agreement with our result  in the next section. Rather than look for power law absorption, it is more common to characterize absorption in glassy media by e.g. its peaks \cite{gilroy1981asymmetric}.
%Peaks in the acoustic absorption are more common to analyze and model, \cite{gilroy1981asymmetric}

	\subsection{Energy distribution}
	
	%%%%%%%%%%%%%%%%%%%%%%%%%%%%%%%%%%%
The Arrhenius relation allows the transformation between the frequency, time and energy  relaxation integrals, \eqref{eq:MultipleRelaxation}, \eqref{eq:MultipleRelaxationTime},  and \eqref{eq:MultipleRelaxationEnergy}. The energy distribution  may be written as:
\begin{equation}
g_E(E_a) = \frac{\Omega}{k_B T} \; g_\Omega(\Omega) =   \frac{\tau}{k_B T} \; g_\tau(\tau),
\label{eq:EnergyTransformations}
\end{equation}
%%%
where the relationship between activation energy and relaxation time is given by \eqref{eq:Arrhenius}. 
In the case of the power-law absorption of interest here, given by \eqref{eq:pdfRelaxationFrequencies} and \eqref{eq:pdfRelaxationTimes}, we have 
\begin{equation}
g_E(E_a) = \frac{K_y}{k_B T} {\Omega^{y-1}} = \frac{K_y {\tau_0^{1-y}}}{k_B T} \;  \E^{-\frac{E_a}{k_BT}(y-1)},
\label{eq:EnergyPdf}
\end{equation}
where $\tau_0$ comes from \eqref{eq:Arrhenius}. This is an exponential distribution, but due to the presence of the $y$, this result is more specific than \eqref{eq:GlassExp}. 

\begin{figure}[tb]
\centering
 \includegraphics[width=\columnwidth]{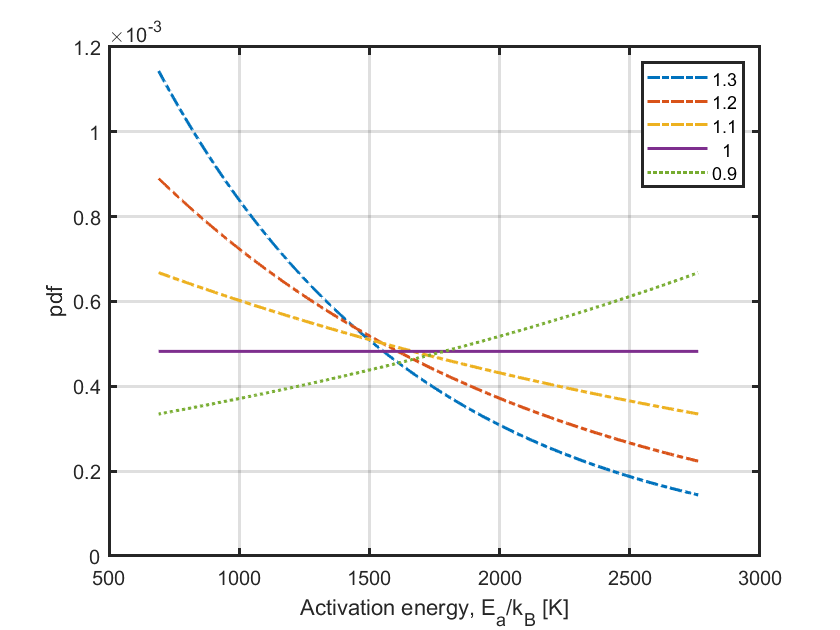}
  \caption{Normalized energy distribution function of \eqref{eq:EnergyPdf} plotted for power-law exponents $y$ in the range from 0.9 to 1.3. The ratio of the upper and lower energy limits is 3 and therefore this plot corresponds to 3 decades of frequency variation.}
  \label{fig:MultipleSlopes}
\end{figure}

The equation is plotted in Fig.~\ref{fig:MultipleSlopes} with the power-law exponent, $y$, as a parameter. The range of activation energies, $E_a/k_B$, is from 691 to 2072 K. Their ratio is 3 and so it corresponds to 3 decades of frequency according to \eqref{eq:Arrhenius}. The lower and upper energy limits correspond to 100 MHz and 100 kHz respectively  for a value of $\tau_0^{-1}=2 \pi \cdot 10^9$.

The most interesting case is the energy distribution corresponding to linearly increasing absorption, $y=1$. Surprisingly, that corresponds  to a flat activation energy distribution. This is not a feature which has  received much attention in the energy landscape interpretation of e.g.~glass. As $y > 1$, the distribution falls off for higher energies and its mean value, assuming that the distribution is defined for all energies from 0 to infinity, is $k_B T/(y-1)$.

The energy distribution can be interpreted as a probability density function, $p(x)$, with proper normalization. The concept of Shannon entropy will aid in understanding its properties:
\begin {equation}
H(p) = - \int_a^b p(x) \ln p(x) \D x.
\end{equation}
The density which maximizes entropy is a uniform or flat probability density function, when $a$ and $b$ are both finite.  This interpretation of the flat energy landscape therefore corresponds to one with maximal randomness or one with the fewest possible constraints. Further, if the entropy is maximized over an infinite interval, $a=0$, $b=\infty$, with the constraint of a given mean value, the result is an exponential probability density function \cite[Table 1]{park2009maximum}. Under that condition, the exponential solution in \eqref{eq:EnergyTransformations} also exhibits maximum randomness.

So far we have given arguments for why \eqref{eq:EnergyPdf} is valid for power-law media where relaxation takes place at the nanometer scale. In the next section the size scale will be expanded.

\section{Biological materials and energy landscapes}

Interestingly, the  energy landscape interpretation is used over a much larger range of scales in \cite[Chap.~1]{wales2004energy}. Wales's book starts by outlining three different fields: 
``The structure and dynamics of atomic and molecular clusters, the folding of proteins, and the complicated phenomenology of glasses are all manifestations of the underlying potential energy surface''

The first and last of these fields have already been mentioned as structural and chemical relaxation and as the effect of structural defects in glasses.
Surprisingly, energy landscapes are used even for proteins and folded proteins, including micelle formation \cite{raschke2019non}.
That would include most, if not all, of the materials mentioned in the Introduction, from the nanometer scale up to the sub-micron range.

\subsection{Validity of the Arrhenius relation}

One thing is that properties are described by an energy landscape, another is the validity of the Arrhenius relation for biological materials. 
As stated in \cite{bi2014energy}, the Arrhenius relation comes from statistical mechanics and is valid for a system which transitions from one metastable state to another.  They argue that although these assumptions are not necessarily valid in biological tissue, analogs to the relevant parameters ``exist in cells and likely govern cell motility."

The energy landscapes of proteins and glasses also have many properties in common. Proteins may in some cases be regarded as two-state systems, such as when  ``an ion channel can be open or closed, a hemoglobin or myoglobin protein can have bound oxygen or not'' \cite{glockle1995fractional}. 

The Arrhenius law is therefore not uncommon to use for characterizing the energy landscapes of cells and proteins as well. Its validity is not as universal as in the fields discussed previously \cite{frauenfelder1991energy}, and there are several unsolved questions in this field. We assume here that the Arrhenius equation may be applied,  although the physical basis is not as solid as for processes at the atomic and molecular scale.

%\cite[p.~68]{wales2004energy}: ``An amorphous material is one that lacks the long-range order of a crystal,'' ``A glass can then be defined as an amorphous solid that exhibits a glass transition, where derivative thermodynamic properties such as the heat capacity change abruptly''

\subsection{Soft glassy materials}

There is independent evidence that human tissue cells under the influence of shear may be modeled with soft glassy rheology  \cite{zhou2009universal}, \cite{kollmannsberger2009active}, \cite{khodadadi2015protein}. This lends support to the just mentioned descriptions of biological materials.
The response often varies with $\omega^\beta$, where $\beta$ is in the range 0.1 to 0.5 \cite{kollmannsberger2009active}, \cite{parker2019towards}. According to results cited in the %Supplement, 
Appendix, this corresponds to an absorption that varies with $y=\beta+1$ and therefore $y$ in this case will fall in the range 1.1 to 1.5.

The properties of soft glasses correspond to those of glasses at temperatures between the glass temperature and the melting point, $T_g < T < T_m$. The cell's mechanical properties are determined by the crowded interior of the cell. This is analogous to what takes place in a colloidal suspension and leads to the complex shear modulus following a weak power law over several frequency decades with a near constant power law exponent, similar to that of Appendix %Supplement 
Eq.~\eqref{eq:PowerLawDynamicModulus}. 
Cells are very soft relative to the materials they are made from, similar to how a wool jumper is soft compared to the wool fibers that it comprises. 
Cells are examples of a disordered metastable material which exists in a state far from thermodynamic equilibrium. %They are characterized by a creep response which is a power law of time which power $\beta$ in the range 0.1 to 0.5. 
The energy landscape is comprised of the binding energy between neighboring proteins. %A deformation due to 
An incoming wave may cause a hop from one of the states mentioned above to the other,
%a hop between energy wells 
where the %deformation 
required energy is taken from the wave, i.e.~leading to heating and absorption of wave energy.

\subsection{Non-Arrhenius behavior}

As mentioned in the previous section, the typical range for $y$ for shear waves in biological tissue is up to $1.5$.  Although $y=1$ is the most common value for compressional waves in medical ultrasound, the value may reach up to $1.5$ even in this field \cite{kadaba1980attenuation}. If the Arrhenius equation is valid, this could mean that the energy distribution  is skewed towards lower energies as in Fig.~\ref{fig:MultipleSlopes}. 

Alternatively, it could also mean that the Arrhenius equation is no longer valid for some biological materials or under specific conditions. 
The stretched exponential  has been proposed as an alternative:
\begin{equation}
\tau = \tau_0 \E^{\left(E_a/k_BT\right)^\gamma}, 
\label{eq:StretchedArrhenius}
\end {equation}
where the range $0 < \gamma < 1$ leads to a stretched exponential, $\gamma>1$ gives  a compressed exponential, and $\gamma=2$ results in a gaussian distribution. 
Repeating the steps leading from \eqref{eq:MultipleRelaxationEnergy} to \eqref{eq:EnergyPdf}, gives a new energy distribution:
\begin{equation}
g_E(E_a)  =  \frac{K_y {\tau_0^{1-y}}}{k_B T} \; \gamma \cdot \E^{- \left( \frac{E_a}{k_BT}\right)^{\gamma} (y-1)}
\cdot \left( \frac{E_a}{k_BT}\right)^{\gamma-1},
\label{eq:PDF-nonArrhenius}
\end{equation}
which for $\gamma=1$ contains \eqref{eq:EnergyPdf} as a special case.

When $\gamma>1$, this equation may predict that the flattest energy distribution  occurs for $y>1$, but the energy distribution is no longer exactly flat as is the case for $\gamma=1$ and $y=1$. As an example $\gamma=1.5$ results in the flattest energy distribution for $y=1.1$. Eq.~\eqref{eq:PDF-nonArrhenius} in combination with a flat energy landscape may therefore only partially explain power-law exponents, $y$, above 1.

%\subsection{Sensitivity to energy distribution in the band-limited case} 
\subsection{Band-limited power laws} 

As is clear from the examples of the Introduction, the  power-laws are only observed over a limited bandwidth. In \cite{nasholm2013model} and \cite{pierce2021acoustic} it is  shown that \eqref{eq:MultipleRelaxation} and \eqref{eq:MultipleRelaxationTime} result in a good fit to \eqref{eq:PowerLawabsorption} even in that case. This means that each  component in the integral  mainly affects frequencies in the vicinity of its relaxation frequency.
Band limiting to a range  from $\Omega_L$ to $\Omega_H$ in \eqref{eq:MultipleRelaxation}, implies that the asymptotes of the power-law absorption will be %\eqref{eq:PowerLawabsorption} will be:
\begin{align}
 \alpha(\omega) \propto
  \begin{cases}
    \omega^2, 	& \omega \ll \Omega_L,\\ 
   \omega^{y}, & \Omega_L \ll \omega \ll \Omega_H,\\ 
   \alpha_\infty,        	& \Omega_H \ll \omega. 
  \end{cases}
  \label{eq:Bandlimited}
\end{align}
A high-frequency limit like that of \eqref{eq:Bandlimited}, corresponding to a lower energy limit, may in fact be required for physical reasons as a passive medium requires that the absorption should not increase faster than  $\omega^1$ as $\omega$ approaches infinity \cite{holm2017restrictions}.
As noted, the low-frequency limit corresponding to an energy distribution which stops at a maximum energy value, is sometimes observed in e.g., sub-bottom sediments \cite{Holm2023lowfrequencylimit}. The exponential relationship of the Arrhenius law also means that a band-limiting corresponds to a relatively narrow range of activation energies.

\begin{figure}[tb]
\centering
 \includegraphics[width=\columnwidth]{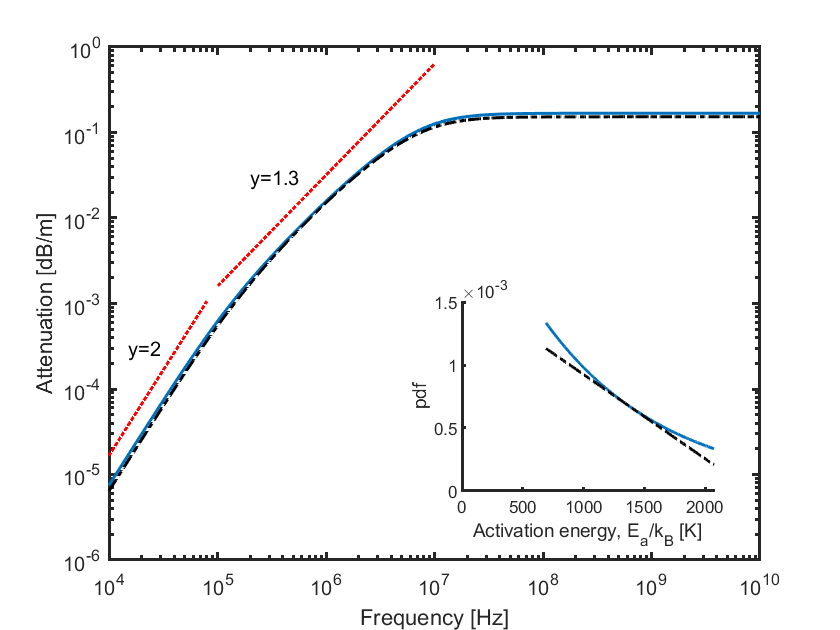}
  \caption{Band-limited example with power-law $y=1.3$. Comparison between exact energy distribution (blue, solid line), and a linear approximation (black, dashed line), and with reference  curves with slopes $y=1.3$ and $y=2$ above them. \\
The inset shows the exact (blue, solid line) and linearized (black, dashed line) energy distributions with $\tau_0^{-1}=2 \pi \cdot 10^8$, $T=300$ K, $f_{L} = 10^5$ and $f_{H}=10^7$ Hz corresponding to $E_a/k_b$ between 2072 and 691 K.}
  \label{fig:Exact-linearized}
\end{figure}

In the band-limited case, the skewed distribution of \eqref{eq:EnergyPdf} may be approximated by the first term in a Taylor series about a point in the middle of the energy range, $E_c$. This point corresponds to a frequency $\Omega_c=\sqrt{\Omega_L \Omega_H}$ and the linear approximation of the distribution is:
\begin{equation}
    g_E(E_a)\approx g_E(E_c) \left[ 1 + \frac{1-y}{T} \frac{E_a-E_c}{k_B}\right],
\end{equation}
which is an acceptable approximation for a narrow frequency range and a power-law exponent near unity. Fig.~\ref{fig:Exact-linearized} shows an example that demonstrates that when \eqref{eq:PowerLawabsorption} is only given over two decades, the exact shape of the energy distribution is not critical. The difference between the absorption for the exact and the approximated linear case, are minor and most likely often smaller than the measurement error. The figure also illustrates the lower and upper asymptotic values given by \eqref{eq:Bandlimited}.

\section{Discussion}

%\subsection{Related results}
Glasses have a universal property at low  temperatures, 0.1 to 10 K, where $Q^{-1}(\omega;T)$ is found to be nearly independent of temperature T as well as  frequency $\omega$ \cite{shukla2022universality}.
The property that elastic wave absorption depends on frequency in a linear way around room temperature is a similar universal property. We have shown here that it results from a flat activation energy distribution. 

An alternative way of deriving an energy distribution similar to \eqref{eq:EnergyPdf} is found in the theory for soft glassy materials \cite{sollich1997rheology}, \cite{sollich1998rheological} 
where the medium was modeled as a Maxwell-Wiechert model as in 
Fig.~\ref{fig:MechanicalModels} and Eq.~\eqref{eq:Maxwell-WiechertAbsorption} of the Appendix. % Supplement.
That theory  gives a material description in the form of a partial differential equation which expresses how regions rearrange to new positions, valid for low frequencies. The main variable is $g_E(E_a)$, the probability  for finding an element  
trapped in a barrier between the two wells of height $E_a$.

The theory contains a constant which is an attempt frequency, and an activation factor on the same form as the Arrhenius equation. 
The theory is expressed in normalized units with a central parameter being the mean-field noise temperature, $y$. A glass transition occurs at $y=y_g=1$ and the material approaches the  fluid state for $y=2$. Another input is a prior distribution of traps which it is argued has an exponential tail, $p_a(E_a) = \exp(-E_a/y_g)$ \cite{bouchaud1992weak}. 

The equilibrium distribution of energies above the glass transition is given by  $g_E(E_a) \propto \E^{E_a/y} p_a(E_a)$, and although the end result is not stated explicitly in \cite[Sect.~IV.A]{sollich1998rheological}, it is an exponential distribution, $g_E(E_a) \propto \E^{-E_a(y-1)/y}$.
%\begin{equation}
%    g_E(E_a) \propto \E^{\frac{-E_a}{y}(y-1)}.
%    \label{eq:SollichPdf}
%\end{equation}
%
Since $y$ is $k_bT/E_0$  \cite{gilroy1981asymmetric}, this expression is analogous to \eqref{eq:EnergyPdf}. The dependence of  $y$ on temperature outlines one way of testing \eqref{eq:EnergyPdf} and new results with ultrasound  properties of tissue at low temperature could potentially be used \cite{liang2024review}. However the possible temperature range is rather limited compared to the range of temperature variation that glasses are tested under. The maximal temperature range that  tissue can be subjected to without structural damage is 5--40  \arcdeg C \cite{peters1997applicability}.

Further it is demonstrated how this model leads to a dynamic modulus ${E}(\omega) = {E}'(\omega) + \I {E}''(\omega)$  where both the real and the imaginary components are proportional to $\omega^{y-1}$, as for the fractional Kelvin-Voigt model of Appendix %Supplement 
Fig.~\ref{fig:FractionalModels} and Eq.~\eqref{eq:PowerLawDynamicModulus}. 

Thus the noise temperature and the fractional order have a simple relationship, $y=\beta + 1$. The soft glassy model therefore provides an interpretation of the fractional order, $\beta$ as well as for $y$.  The special case of concern in this article is found in the limit, as $y$ approaches one, i.e. the material approaches the glass temperature. 
It is also evident that the soft glassy model result of \cite{sollich1998rheological}
%\eqref{eq:SollichPdf}
for the energy distribution resembles our result \eqref{eq:EnergyPdf}. In the soft glassy theory it is found in a bottom-up way, while our independent derivation is more of a top-down approach.

A philosophically inclined reader may have noted that we have not followed the often-desired hypothetico-deductive way of arguing strictly, as so far, 
only the temperature dependence of $y$ has been proposed as a way of partially testing our hypothesis.
%no new testable prediction can be deduced from our hypothesis. Instead, we have used 
The main argument here is 
inference to best explanation, or abductive reasoning, arguing that an Arrhenius-type relation and a formulation in terms of an energy landscape unite many of the materials that display linearly increasing absorption, even those where other processes than thermal activation at the molecular level take place. Additionally, a flat energy landscape with maximal randomness makes sense as it can be interpreted as a maximum entropy distribution of energies.

\section{Conclusion}

It is remarkable that elastic wave absorption depends on frequency in a linear way around room temperature universally across applications as diverse as seismology, seismics, subbottom acoustics, and medical ultrasound. 

Such absorption is the result of a large number of relaxation processes, expressed by a weighted  relaxation integral over frequency or over time. 
It is however hard to argue physically why the particular weighting that gives rise to power-law absorption, should occur. We have argued here, based on properties of atomic and molecular clusters, proteins, and glasses that an energy landscape formulation is fundamental. Further the Arrhenius expression for activation energy, despite being formally derived in statistical mechanics, is often valid. This enables the transformation of the multiple relaxation formulation to an integral over energies in an energy landscape. %The formulation derived here is not common in e.g.~ultrasonics and seismics. 

A  macroscopic property, absorption, is  linked to properties at the mesoscale level, the shape of the energy landscape.
The interesting case of linearly increasing absorption corresponds to a flat activation energy distribution, i.e., all energies are equally probable. A flat energy distribution indicates a form of equilibrium, and properties of the energy landscape may enable a deeper understanding of both the conditions for linearly increasing absorption with frequency as well as the origin of power-law relaxation responses in general.

%\section*{Acknowledgments}
\section*{Acknowledgments}   %% 22.05.2024 changed spelling
We want to thank Svein-Erik Måsøy and Kevin Parker for valuable discussions of the topic of this paper, and to Yuri Galperin, Sven Peter N\"asholm and Ralph Sinkus for comments on an early version of the manuscript.
%\end{acknowledgments}

%%%%%%%%%%%%%%%%%%%%%%%%%%%%%%%%%%%%%%%%
%\newpage
\section*{Appendix}
%\beginsupplement
%\appendix
%\section{Alternative models of absorption}

%\renewcommand{\theequation}{S.\arabic{equation}}
%\setcounter{equation}{0}

There are ways of modeling power-law absorption that are different from the multiple relaxation model. They are referred to in some of the references of the main text and therefore two alternatives are discussed here: A mechanical model consisting of an infinite network of springs and dampers (Maxwell-Wiechert model), and a fractional Kelvin-Voigt model. %and a fractional Zener model. 

\subsection{Mechanical  interpretation}

\begin{figure}[b]
\centering
 \includegraphics[width=0.6\columnwidth]{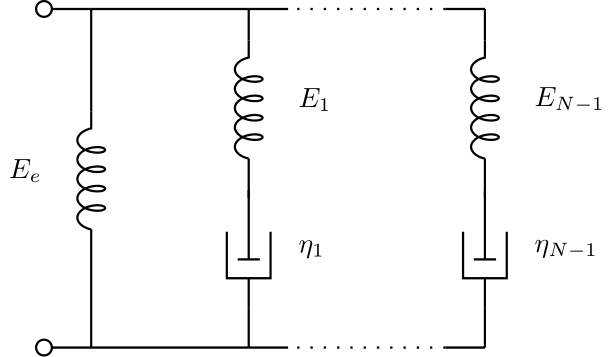}
  \caption{Maxwell-Wiechert model with $E$'s denoting modulus of elasticity and $\eta$'s  viscosity.}
  \label{fig:MechanicalModels}
\end{figure}
The multiple relaxation models can be given a mechanical interpretation by considering the discrete Maxwell-Wiechert model consisting of parallel branches of springs and dampers in series, as shown in Fig.~\ref{fig:MechanicalModels}. Its dynamic modulus is:
% Eq 3.54 in \cite{holm2019waves}:
\begin{equation}
    {E}(\omega) = E_e +  \sum_{n=1}^{N-1} \frac{E_n \; \I \omega \tau_n}{1+\I \omega \tau_n}
     \label{eq:Dicrete-Maxwell-Wiechert}
\end{equation}
where $\tau_n = \eta_n/E_n$. Note that in this appendix $E$ means elasticity modulus and not energy as in the main text. The spring represented by $E_e$, the equilibrium value, 
is required in order to make this a model for a solid. In the limit, it is a continuous model 
where $\hat{E}(\tau)$ is the relaxation spectrum or distribution of elastic moduli \cite[Sect.~4.1.1]{tschoegl1989phenomenological}, whether they are shear or bulk moduli:
\begin{equation}
    {E}(\omega) =   E_e +  \int_0^\infty \frac{\hat{E}(\tau) \; \I \omega }{1+\I \omega \tau} \; \D \tau,
     \label{eq:Maxwell-Wiechert}
\end{equation}

There is a direct relation to the absorption of 
\eqref{eq:MultipleRelaxation} and \eqref{eq:MultipleRelaxationTime} 
which can be found from linearized conservation of momentum and energy. In that case the dispersion relation can be shown to be \cite[Sect.~3.5]{holm2019waves}:
\begin{equation}
    k^2(\omega) = \rho_0 \frac{\omega^2}{E(\omega)}
\end{equation}
Absorption is found from the imaginary part of the wavenumber $k(\omega)$: %= -\Im{k(\omega)} 
\begin{equation}
    \alpha(\omega) = - \sqrt{\frac{\rho_0}{E_e} } \omega \;  \Im \left[ 1 + \int_0^\infty \frac{\I \omega \; \hat{E}(\tau)/E_e }{1+\I \omega \tau} \; \D \tau \right]^{-1/2}
\end{equation}

Assuming that $\hat{E}(\tau) \ll E_e$, i.e., all loss mechanisms are weak, the absorption is:
\begin{equation}
    \alpha(\omega) \approx \sqrt{\frac{\rho_0 E_e}{2}} \; \omega^2 \int_0^\infty \frac{\hat{E}(\tau) }{1+\omega^2 \tau^2} \; \D \tau 
    \label{eq:Maxwell-WiechertAbsorption}
\end{equation}
which is similar to \eqref{eq:MultipleRelaxationTime} with $\hat{E}(\tau)  \propto g_\tau(\tau) \tau$. The exact expression for  $\hat{E}(\tau)$ in the power-law case is:
\begin{equation}
\hat{E}(\tau)  = \sqrt{\frac{2}{\rho_0 E_e}} \;  K_y \tau^{-y}.
\end{equation}
This description therefore gives insight into how the relaxation integrals of \eqref{eq:MultipleRelaxation} and \eqref{eq:MultipleRelaxationTime} 
can be realized in terms of elementary mechanical models, but it does not give us much insight into why  $\hat{E}(\tau)$ is shaped in a particular way in the case of power-law absorption.

%\begin{equation}
%g_V(V) = \Omega \; g_\Omega(\Omega) = \tau g_\tau(\tau) = \hat{E}(\tau)  = \Omega^{y-1}.
%\label{eq:EnergyPdf2}
%\end{equation}

\subsection{Fractional Kelvin model}

\begin{figure}[t]
\centering
 \includegraphics[width=0.35\columnwidth]{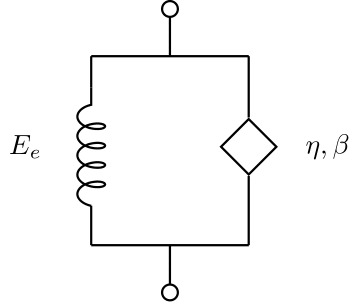}
  \caption{Fractional Kelvin-Voigt medium model with $E_e$ denoting modulus of elasticity and $\eta$ denoting pseudo-viscosity of order $\beta$.}
  \label{fig:FractionalModels}
\end{figure}

Fractional viscoelasticity is an alternative  way of expressing power-law absorption. Its advantage is that it requires a small number of parameters. 

In particular the fractional Kelvin-Voigt mechanical model shown in Fig.~\ref{fig:FractionalModels} has been used for tissue discrimination in elastography, i.e.~shear wave imaging in tissue \cite{sinkus2018rheological}, and has been recommended for general use in elastography \cite{parker2019towards}. The model consists of a fractional damper of order $\beta$ in parallel with a spring \cite[Sect.~3.1]{Mainardi2010}, \cite[Sect.~5.2]{holm2019waves}, where the stress is given by 
\begin{equation}
    \sigma(t) = E_e \varepsilon(t) + \eta \; \frac{\partial^\beta \varepsilon(t)}{\partial t^\beta},    
\end{equation}
where $\varepsilon(t)$ is strain, $E_e$ is the spring's modulus of elasticity, and $\eta$ is a pseudo-viscosity in units Pa/s$^\beta$. The dynamic modulus, the Fourier transform of the stress impulse response, is 
\begin{equation}
 {E}(\omega) = E_e + \eta \; (\I \omega)^{\beta}. 
 \label{eq:PowerLawDynamicModulus}
\end{equation}
Here both the real and the imaginary parts have components that are proportional to $\omega^\beta$. 
It is  shown in \cite[Sect.~5.6]{holm2019waves} that the wave equation in this case has a solution which gives rise to an attenuation which follows $\omega^{\beta + 1}$ for low frequencies. Thus the power-law absorption of \eqref{eq:PowerLawabsorption} for this model is: % $y=\beta+1$ for this model, which is used in \cite{sollich1998rheological}.
\begin{equation}
\alpha(\omega) = \alpha_0\omega^{y} =\alpha_0\omega^{\beta+1}.
\end{equation}

\section*{Author Declarations}

\subsection*{Conflict of Interest Statement}
The authors have no conflicts to disclose.

\subsection*{Data Availability}
No data was collected for this research.

\subsection*{Ethics Approval}
No ethics approval was required for this research.

%\bibliography{references}%

%%%%%%%%%%%%%%%%%% from bbl-file
%

%%%%%%%%%%%%%%%%%%

\end{document}